\documentclass[graybox, envcountchap]{svmult}

\usepackage{mathptmx}        
\usepackage{amsmath}
\usepackage{amssymb}
\usepackage{color}
\usepackage{helvet}          
\usepackage{courier}         
\usepackage{dirtree}

\usepackage{makeidx}        
\usepackage{graphicx}        
\usepackage{subfig}

\usepackage{multicol}        
\usepackage[bottom]{footmisc}

\usepackage{hyperref}        
\hypersetup{colorlinks=true,urlcolor=blue}

\usepackage[misc]{ifsym}

\makeindex             

\begin{document}


\title{Wind from the Hot Accretion Flow and Super-Eddington Accretion Flow}
\author{Hai Yang and Feng Yuan}
\institute{Hai Yang (\Letter) \at Tsung-Dao Lee Institute, Shanghai Jiao Tong University, 1 Lisuo Road, Pudong New Area, Shanghai, 201210, China \email{hai-yang@sjtu.edu.cn}
\and 
Feng Yuan \at Center for Astronomy and Astrophysics and Department of Physics, Fudan University, Shanghai 200438, China \email{fyuan@fudan.edu.cn}}
%
%
\maketitle

\abstract{ 
Wind is believed to be widespread in various black hole accretion flows. However, unlike the wind from thin disks, which have substantial observational evidence, the wind from hot accretion flows is difficult to observe due to the extremely high temperatures causing the gas to be almost fully ionized. Its existence was controversial until recent theoretical work demonstrated its presence and strength, which was subsequently confirmed by observations. Although there have been some new observations recently, the main progress still comes from theoretical studies. These studies investigate the effects of different magnetic fields and black hole spins on the wind, providing insights into properties such as mass flux and wind velocity. Wind is typically produced locally within the Bondi radius, and even wind generated on a small scale can propagate far beyond this radius. The situation with super-Eddington wind is similar, despite some recent observations, the main advances rely on theoretical studies. Recent research comparing the momentum and energy of wind and jets suggests that wind plays a more crucial role in active galactic nuclei feedback than jets, whether the wind originates from hot accretion flows or super-Eddington accretion flows.}



\section{Introduction}
\label{sec:1}

Black hole accretion is a prevalent and fundamental high-energy astrophysical phenomenon in the Universe and is generally regarded as the primary energy source for many observed phenomena, such as active galactic nuclei (AGNs), black hole X-ray binaries (BHBs), and gamma-ray bursts. Depending on the temperature of the accretion flow, black hole accretion is commonly categorized into two modes: cold and hot modes. These classification criteria are based on the mass accretion rate of the black hole. In the cold mode, the accretion rate exceeds approximately $2\%\,\dot{M}_{\rm Edd}$, where $\dot{M}_{\rm Edd}=10\,L_{\rm Edd}/c^{2}$ represents the Eddington accretion rate and $L_{\rm Edd}$ is the Eddington luminosity. Conversely, in the hot mode, the accretion rate is lower than about $2\%\,\dot{M}_{\rm Edd}$\cite{Yuan:2014}.

In cold accretion modes, the accretion flow is optically thick, making radiation a crucial factor. Conversely, hot modes are typically optically thin, resulting in radiation being of lesser importance. Cold mode further segregates into two accretion disk models. Firstly, there is the standard thin disk \cite{Shakura:1973,Pringle:1981} characterized by accretion rates lower than $\dot{M}_{\rm Edd}$. This disk is geometrically thin, emitting thermal blackbody-like radiation that successfully describes luminous AGNs \cite{Pringle:1981, Koratkar:1999, Abramowicz:2013} and BHBs in their thermal state\cite{McClintock:2014}.

Secondly, there's the super-Eddington accretion flow\cite{Abramowicz:1988, Ohsuga:2005, Jiang:2014, Sadowski:2014, McKinney:2014} characterized by accretion rates surpassing $\dot{M}_{\rm Edd}$. In this scenario, the accretion disk is geometrically thick, with the gas being too optically thick to radiate all dissipated energy locally. Consequently, radiation becomes trapped and inwardly advected by the accretion flow. This model is employed to elucidate certain observations like tidal disruption events (TDEs)\cite{Dai:2018}, SS433\cite{Fabrika:2004}, and ultraluminous X-ray sources\cite{Watarai:2001}.

The hot modes usually correspond to an advection-dominated accretion flow (ADAF), also known as hot accretion flow \cite{Narayan:1994, Yuan:2014}. In this scenario, the accretion disk is geometrically thick and the gas is optically thin, meaning that energy is accreted into the black hole before it can be cooled by radiation. This model has been applied to a wide variety of accretion systems, including the supermassive black holes (SMBHs) Sagittarius A* (Sgr A*), M87, low-luminosity AGNs, which are thought to be prevalent in most galaxies, and the hard/quiescent states of black hole binaries.

\section{Theoretical study of wind from hot accretion flows}
\label{sec:2}

AGN feedback is thought to play a key role in the formation and evolution of galaxies\cite{Fabian:2012, Kormendy:2013, Naab:2017}. The wind from hot accretion flow may be one of the most important feedback media\cite{Weinberger:2017, Yuan:2018, Yoon:2019}. Recent IllustrisTNG
cosmological simulations\cite{Weinberger:2017} research has found that winds from hot accretion flow are needed to solve some problems encountered in galaxy formation, such as the interaction of the wind with the interstellar medium at the galaxy scale is required to reduce the star formation efficiency in the most massive halos. Yuan et al.\cite{Yuan:2018} studied the feedback effects involving wind and radiation from AGNs and found that wind plays a dominant role in controlling star formation and black hole mass growth.

In this section, we primarily introduce the latest research on winds in hot accretion flows. For a more comprehensive review of hot accretion flows and the driving mechanisms of winds from hot accretion flows, refer to \cite{Yuan:2014}.

The study of winds in accretion flows has long been a crucial research topic in this field and represents significant progress. Its importance and attraction stem mainly from two aspects. Firstly, winds constitute a fundamental component of black hole accretion flows. Research indicates that a substantial portion of gas flowing in from large radii is not accreted by the black hole but instead is predominantly converted into wind. Such powerful winds profoundly impact the dynamic properties of the accretion flow, including angular momentum transfer, mass accretion rate, temperature, density, and consequently influence radiation emitted from the accretion flow.

Secondly, AGN feedback is believed to play a pivotal role in the formation and evolution of galaxies \cite{Fabian:2012, Kormendy:2013, Naab:2017}. Winds from hot accretion flows may constitute one of the most significant feedback mechanisms \cite{Weinberger:2017, Yuan:2018, Yoon:2019}. Recent cosmological simulations with IllustrisTNG \cite{Weinberger:2017} have revealed that winds originating from hot accretion flows are crucial for addressing certain challenges in galaxy formation. For instance, the interaction of these winds with the interstellar medium on galactic scales is necessary to reduce the star formation efficiency in the most massive halos. Yuan et al. \cite{Yuan:2018} investigated feedback effects involving winds and radiation from AGNs, highlighting winds' dominant role in regulating star formation and the growth of black hole masses.

Narayan's earliest analytical self-similar solution to hot accretion flows, as proposed in \cite{Narayan:1994}, suggested the presence of strong winds. In 1999, a milestone in wind studies was achieved when Stone et al. \cite{Stone:1999} conducted the first comprehensive global numerical simulation of hot accretion flows. Their study calculated the inflow, outflow, and net accretion rate of mass flux.
\begin{eqnarray}
\dot{M}_{\rm in}  = -2\pi r^2\int^{\pi}_{0} \rho \rm{min}(v_r,0)\sin{\theta}d\theta 
\label{eq:01}
\end{eqnarray}

\begin{eqnarray}
\dot{M}_{\rm out}  = 2\pi r^2\int^{\pi}_{0} \rho \rm{max}(v_r,0)\sin{\theta}d\theta 
\label{eq:02}
\end{eqnarray}

\begin{eqnarray}
\dot{M}_{\rm net}  = \dot{M}_{\rm in} - \dot{M}_{\rm out}
\label{eq:03}
\end{eqnarray}

\begin{figure}[b]

\centering{\includegraphics[scale=.5]{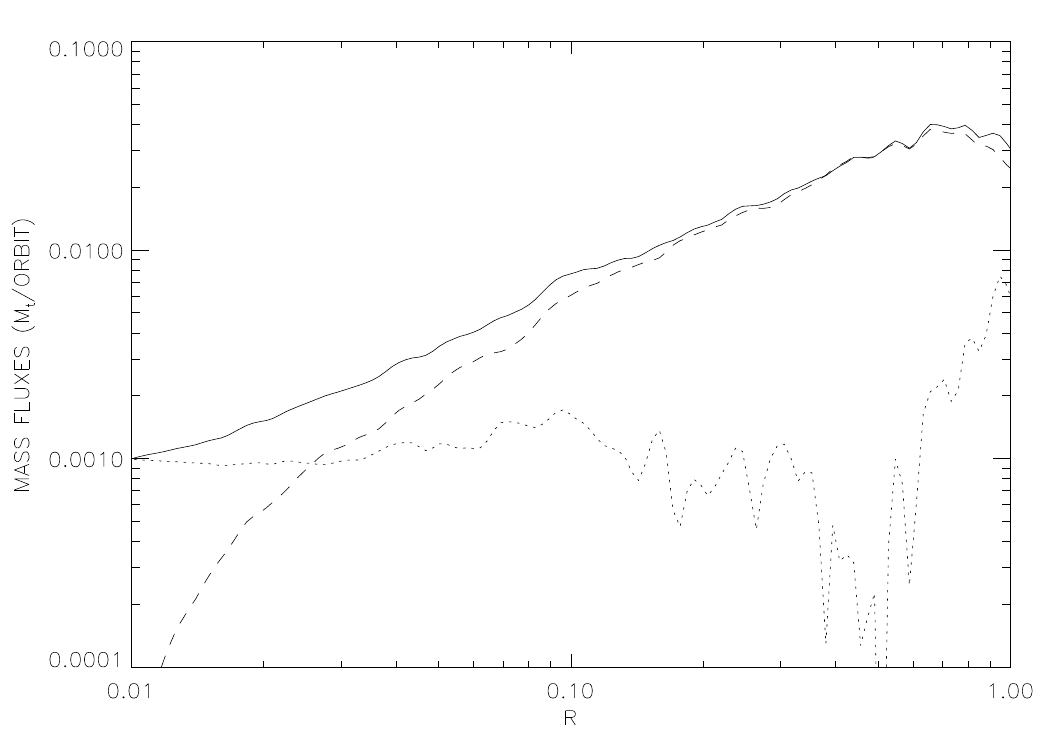}}
%
%
\caption{Radial distribution of different mass accretion rates. The solid, dashed, and dotted lines represent inflow, outflow, and net accretion flow, respectively. Figure adapted from Stone et al. \cite{Stone:1999}.}
\label{fig:1}       
\end{figure}

As shown in Figure \ref{fig:1}, Stone et al. \cite{Stone:1999} found that both the inflow rate $\dot{M}_{\rm in}$ and outflow rate $\dot{M}_{\rm out}$ decrease as the radius decreases, following a power-law relationship,
\begin{eqnarray}
\dot{M}_{\rm in}(r)  = \dot{M}_{\rm in}(r_{\rm out}) (\frac{r}{r_{\rm out}})^{s}, \ \ \ s >0
\label{eq:04}
\end{eqnarray}
where $\dot{M}_{\rm in}(r_{\rm out})$ is the mass flux of inflow at outer boundary $r_{\rm out}$. This result is surprising because it is generally believed that the accretion rate remains constant and does not vary with radius. Early analytical studies often assumed a constant accretion rate independent of radius. The radial distribution of density typically follows $\rho \propto r^{-3/2}$ \cite{Narayan:1994}. However, subsequent numerical simulations, both hydrodynamic (HD) \cite{Stone:1999,Yuan:2010,Yuan:2012b,Li:2013} and magnetohydrodynamic (MHD) \cite{Stone:2001,Hawley:2001,Igumenshchev:2003,Pen:2003,Kato:2004,Pang:2011,Yuan:2012a,McKinney:2012,Narayan:2012}, have shown that the mass accretion rate decreases towards the black hole.

Due to technical constraints, most simulations only achieve equilibrium over two orders of magnitude in radius, potentially affected by boundary conditions and thus reliability. Yuan et al. \cite{Yuan:2012a} extended the equilibrium range to four orders of magnitude using a "two-zone" approach, from approximately $2\,r_{\rm g}$ to $80,000\,r_{\rm g}$, where $r_{\rm g}=GM/c^2$ is the gravitational radius. Their work supports the earlier finding of a power-law distribution of inflow rate and density. They consolidated their findings with previous HD and MHD simulations, whether 2D or 3D, demonstrating a consistent density distribution $\rho \propto r^{-(0.5-1)}$. Variations in the exponent may arise from differences in parameters such as $\alpha$, magnetic field strength, gravitational potential of the black hole, and initial simulation conditions.

To explain the radial variations in inflow and outflow observed in numerical simulations, two main competing models have been proposed. One is the ADIOS (Adiabatic Inflow-Outflow Solution) model \cite{Blandford:1999, Blandford:2004, Begelman:2012} proposed by Blandford et al. This model posits that the presence of wind causes a decrease in the inward accretion rate $\dot{M}_{\rm in}$. However, it does not specify the mechanism driving the wind, assuming instead that some mechanism extracts energy from the accretion flow to generate the wind. Self-similar solutions were derived for 1D and 2D cases, showing an index $0 < s < 1$.

Another model is the CDAF (Convection-Dominated Accretion Flow) model \cite{Narayan:2000, Quataert:2000, Abramowicz:2002, Igumenshchev:2002} proposed by Narayan et al., which is based on the assumption that hot accretion flows are convectively unstable \cite{Narayan:1994}. In the CDAF model, inward angular momentum transfer by convection balances outward transfer by viscous stresses, leading to the cancellation of these effects. This results in convective eddies around the black hole where gas continually convects inward and outward. The gas trapped in these eddies does not accrete onto the black hole, causing an inward decrease in the accretion rate.

The debate over which model is correct has persisted for over a decade and remains unresolved. Resolving this debate hinges on determining the extent to which the outflow in Equation \ref{eq:02} represents actual outflow. Addressing this, Yuan et al. \cite{Yuan:2012a, Yuan:2012b} conducted MHD numerical simulations. In Equation \ref{eq:02}, any $v_{r}>0$ denotes outflow; however, in accretion flows, both real outflow and the outward component of convective eddies contribute. Yuan et al. \cite{Yuan:2012b} thus compared various physical properties of inflows and outflows, including temperature, angular momentum, and velocity. They found systematic and significant differences between inflows and outflows. Analysis of angular momentum distribution in the $\theta$ direction indicated weaker inflow angular momentum where outflow was stronger. In HD simulations, outflow temperatures were consistently higher than inflow temperatures, suggesting buoyancy-driven by convective instabilities in HD accretion flows. In MHD simulations, outflow angular momentum approached Keplerian levels near the equatorial plane, contrasting sharply with low inflow angular momentum. These differences challenge the CDAF model's ability to explain these observations and provide strong evidence that both inflow and outflow mass fluxes are, if not dominant, at least significant.

\begin{figure}[htpb]
\centerline{\includegraphics[scale=.5]{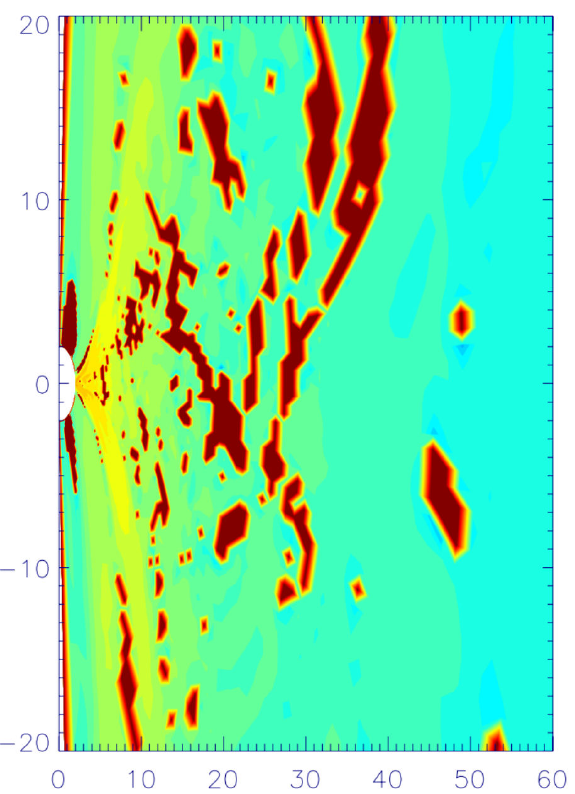}}
\caption{Results of convective stability analysis of MHD accretion flow. Regions colored in red indicate instability. Adapted from Yuan et al. \cite{Yuan:2012b}.}
\label{fig:2}      
\end{figure}

Additionally, they examined the convective stability of accretion flows, finding that while HD flows are convectively unstable, the addition of a magnetic field stabilizes convection. Figure~\ref{fig:2} illustrates these findings, showing that much of the accretion flow areas are not red, indicating convective stability in MHD flows. This result corroborates findings by Narayan et al. \cite{Narayan:2012}.

However, the observed inward decrease in accretion rate $\dot{M}_{\rm in}$ is evident in both HD and MHD simulations, suggesting the CDAF model's inadequacy in explaining these phenomena. Consequently, these results strongly support the existence of winds, aligning more closely with the ADIOS model.

While the aforementioned studies indicate the presence of winds, none quantifies the true strength of these winds—specifically, the mass flux of the wind. The method outlined in Equation \ref{eq:02} for calculating outflow inevitably overestimates the magnitude of the wind mass flow due to its inclusion of turbulent components. Gas may momentarily move outward only to reverse direction and flow inward moments later, limiting accurate measurement to an upper bound. To address this issue, a method to account for turbulence effects is necessary.

Narayan et al. \cite{Narayan:2012} proposed the use of a time-averaged "streamline method" to determine the real outflow. Initially, they conducted time and azimuthal averaging of pertinent physical quantities—notably density $\rho$, velocity $v_{r}$, and $\rho u_{\rm t}$, where $v_{i}$ ($i =r,\, \theta, \, \varphi$)and $u_{\nu}$ ($\nu =t, \, r,\, \theta, \, \varphi$) represent the 3-velocity and 4-velocity of the fluid respectively. This averaging effectively cancels out turbulent fluctuations within the flow's equilibrium region, facilitating a focus on its averaged properties. They defined $\langle v_{r}\rangle _{t\varphi} >0$ as the criterion for genuine outflow and the opposite as inflow.

The outflow mass flux is then calculated using the formula,
\begin{eqnarray}
\dot{M}_{\rm out}  = 2\pi r^2\int^{\pi}_{0}  \rm{max}( \langle \rho v_r \rangle_{t \varphi} ,0)\sin{\theta}d\theta 
\label{eq:05}
\end{eqnarray}

Although this approach effectively removes the effects of turbulence-induced oscillations, it tends to underestimate the mass flux of real outflows. This is because outflows are inherently intermittent and can wander in three-dimensional space. Even at the same spatial location, the velocity $v_{r}$ in the accretion flow may be positive (indicating wind) at one moment and negative (indicating inflow) at the next. Therefore, if a time average is applied first, real outflows may be averaged out or diminished. Consequently, using $\langle v_{r}\rangle _{t\varphi} >0$ as the criterion to identify outflows risks underestimating their true magnitude.

Secondly, they used the Bernoulli parameter as a necessary condition for identifying real outflows, specifically requiring $Be>0$. The Bernoulli parameter typically includes the sum of gravitational potential energy, kinetic energy, and enthalpy. When a magnetic field is present, its contribution must also be considered. In General Relativistic Magnetohydrodynamic (GRMHD) accretion flows, the Bernoulli parameter can be calculated using the following formula \cite{Penna:2013},

\begin{eqnarray}
Be  = - \frac{\langle \rho u_{t} \rangle + \Gamma \langle u u_{t} \rangle + \langle b^{2} u_{t} \rangle}{\langle \rho  \rangle } -1
\label{eq:06}
\end{eqnarray}

where $\langle ... \rangle$ denotes averaging over time and azimuth, $u$ is the internal energy of the gas, $b^{\nu}$ is the four-vector of the fluid frame magnetic field. There is also research utilizing the parameter $\mu$\cite{Narayan:2012,Sadowski:2013}, which is given by,
\begin{eqnarray}
\mu  =  \frac{\langle  T^{p}_{t} \rangle }{\langle \rho u^{p}  \rangle } -1
\label{eq:07}
\end{eqnarray}

In a steady-state axisymmetric ideal MHD flow, $\mu$ is conserved along the streamlines, where the index $p$ represents "poloidal", $T^{\alpha}_{\beta}$ is the stress-energy tensor. Therefore, $\mu$ was used as a criterion for identifying real outflows, indicating that the gas can escape to infinity. This criterion helps to filter out a significant amount of turbulent outflow to a certain extent. However, in a non-steady-state accretion flow, the Bernoulli parameter is not constant along streamlines and may increase outward. This implies that at a certain radius, even if $ Be < 0 $, gas can still escape to infinity.

\begin{figure}[htpb]
\centerline{\includegraphics[scale=.5]{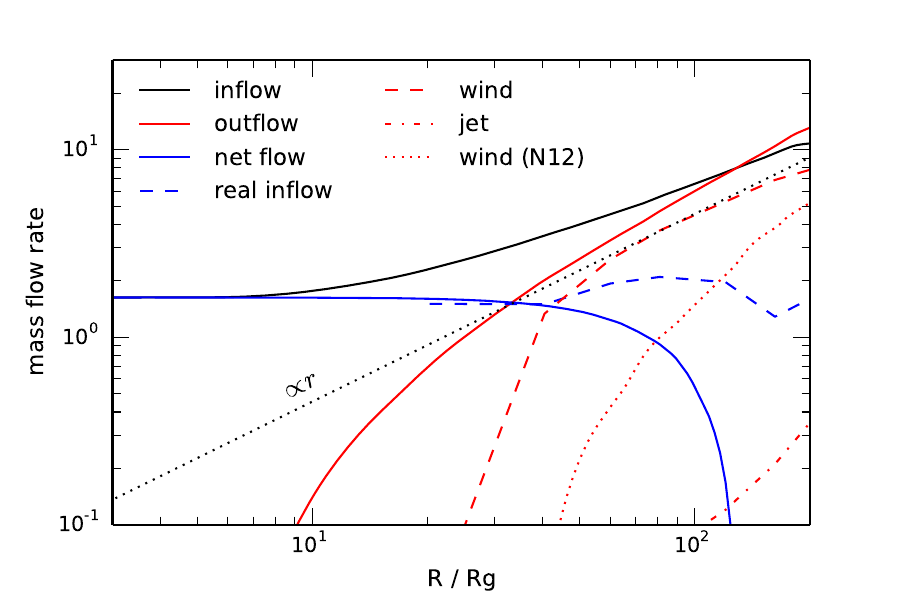}}
\caption{Radial distribution of different mass accretion rates. The black, red, and blue solid lines represent the total inflow, outflow, and net flow calculated based on Equations \ref{eq:01}-\ref{eq:03}, respectively. The red and blue dashed lines represent the real outflow and inflow, respectively, using the 'trajectory method'. The red-dotted line indicates the mass flux of the disk-jet, while the red-dashed line shows the actual outflow obtained from the 'streamline method'. Adapted from Yuan et al. \cite{Yuan:2015}. }
\label{fig:3}      
\end{figure}

Indeed, Yuan et al. \cite{Yuan:2015} found that the Bernoulli parameter $Be$ of the wind can take any value—positive or negative—at any radius. Consequently, $Be$ is not a reliable criterion for real outflow in accretion flows. Using $Be$ as a criterion is likely to filter out some real outflows, thus causing an underestimation of the outflow mass flux. This explains why Narayan et al. \cite{Narayan:2012} found the wind to be very weak, as shown in Figure \ref{fig:3}, and only becoming significant beyond approximately $100 \, r_{\rm g}$.

Therefore, the actual wind mass flux should be between the red solid line (calculated by Equation \ref{eq:02}) and the red dotted line (calculated by Equation \ref{eq:05}) in Figure \ref{fig:3}.

To determine the real wind mass flux, Yuan et al. \cite{Yuan:2015} introduced the ``trajectory" method in the analysis of accretion flows. Trajectories relate to the Lagrangian description of fluid motion, obtained by tracking the continuous motion of fluid elements over time. In contrast, streamlines relate to the Eulerian description of the fluid and are derived by connecting the velocity fields of adjacent fluid elements at a given moment. Although calculating trajectories is more complex than calculating streamlines, it provides a more realistic depiction of fluid element movements.

To obtain the trajectory line, a series of "test particles" are placed at a certain radius in space at a given time. These test particles are not real particles but represent spatial coordinate sets that serve as the starting points for trajectory lines. Their velocities are then obtained through interpolation, allowing the determination of their positions at the next time step $ t + \delta t $. This process continues until the simulation ends or the test particle exits the simulation area. 

A critical aspect of this calculation is ensuring that $ \delta t $ is sufficiently small to achieve convergence, depending on the rate at which the particle changes its speed as it moves. Typically, the time scale for $\delta t$ is the Keplerian timescale.

\begin{figure}[htpb]
\centerline{\includegraphics[scale=.5]{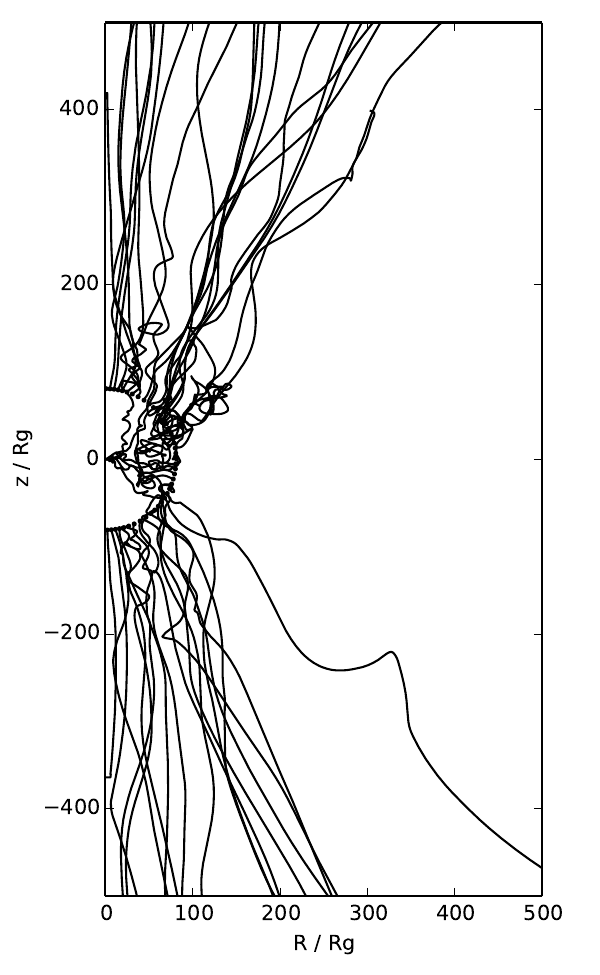}}
\caption{The solid lines in the figure depict trajectories of various gas elements, with (0,0) indicating the location of the black hole. Adapted from Yuan et al. \cite{Yuan:2015}.}
\label{fig:4}      
\end{figure}

Based on the trajectory method, Yuan et al. \cite{Yuan:2015} analyzed the simulation data of 3D GRMHD. The trajectories of some of the test particles are shown in Figure \ref{fig:4}. These trajectories reveal the existence of significant real outflows in the coronal region, providing direct evidence of real outflows rather than relying solely on indirect evidence \cite{Yuan:2012b}. 

To perform this analysis, test particles were first placed at different $\theta$ and $\varphi$ angles with a fixed radius, and their trajectories were obtained. Next, particles that maintained continuous outward motion and did not cross the radius $r$ were identified as real outflows. Finally, the mass flux of the particles identified as real outflow was summed to determine the ``real outflow rate",
\begin{eqnarray}
\dot{M}_{\rm wind}(r)  = \sum_{i} \rho_{i}(r)v_{r,i}(r) r^2 \sin{\theta}\delta \theta_{i} \delta \varphi_{i}
\label{eq:08}
\end{eqnarray}
where $\rho_{i}(r)$ and $v_{r, i}(r)$ are the mass density and radial velocity of the test particle $i$ at the initial position, respectively. $\delta \theta_{i}$ and $\delta \varphi_{i}$ represent the ranges of $\theta$ and $\varphi$ occupied by the test particle $i$, respectively. 
From Equation \ref{eq:08}, we can derive the real outflow mass flux, which is shown as the red dashed line in Figure \ref{fig:3}. It is evident that the real outflow mass flux obtained using the trajectory method (Equation \ref{eq:08}) is significantly larger than the result obtained using the streamline method (Equation \ref{eq:05}). The mass flux of the real outflow is comparable to the accretion rate at the black hole horizon, $\dot{M}_{\rm BH}$, at approximately $40 \, r_{\rm g}$. In contrast, Narayan et al. \cite{Narayan:2012} found this value to be around $100 \, r_{\rm g}$.

Additionally, they found that the mass flux of the wind can be well expressed by:
\begin{eqnarray}
\dot{M}_{\rm wind}(r)  \approx \dot{M}_{\rm BH}(\frac{r}{20\,r_{\rm s}})^s, \ s\approx 1,
\label{eq:09}
\end{eqnarray}

where $ r_{\rm s} \equiv 2 \, r_{\rm g} $. Furthermore, the power-law index $ s $ in Equation \ref{eq:09} aligns closely with the values obtained from the analytical work of Begelman et al. \cite{Begelman:2012}. They also calculated the mass flux-weighted poloidal velocity ($ v_{p} = \sqrt{v_{r}^2 + v_{\theta}^2} $) of the wind and found that it can be well expressed by the following formula,
\begin{eqnarray}
v_{p,\rm wind}(r)  \approx 0.21\,v_{k}(r),
\label{eq:10}
\end{eqnarray}

where $ v_{k}(r) \equiv \sqrt{GM/r} $ is the Keplerian velocity at radius $r$. It should be noted that this formula indicates the poloidal velocity of the wind at radius $ r $. Considering that the wind is constantly generated locally, this is more likely to reflect the poloidal velocity of the wind generated at a radius close to $ r $. Additionally, they found that the poloidal velocity of the test particle was either constant along the trajectory or slightly accelerated outward. Therefore, they provided the terminal poloidal velocity of the wind originating at radius $ r $,
\begin{eqnarray}
v_{p,\rm term}(r)  \approx (0.2 \sim  0.4)\,v_{\rm k}(r),
\label{eq:11}
\end{eqnarray}

\begin{figure}[htpb]
\centerline{\includegraphics[scale=.5]{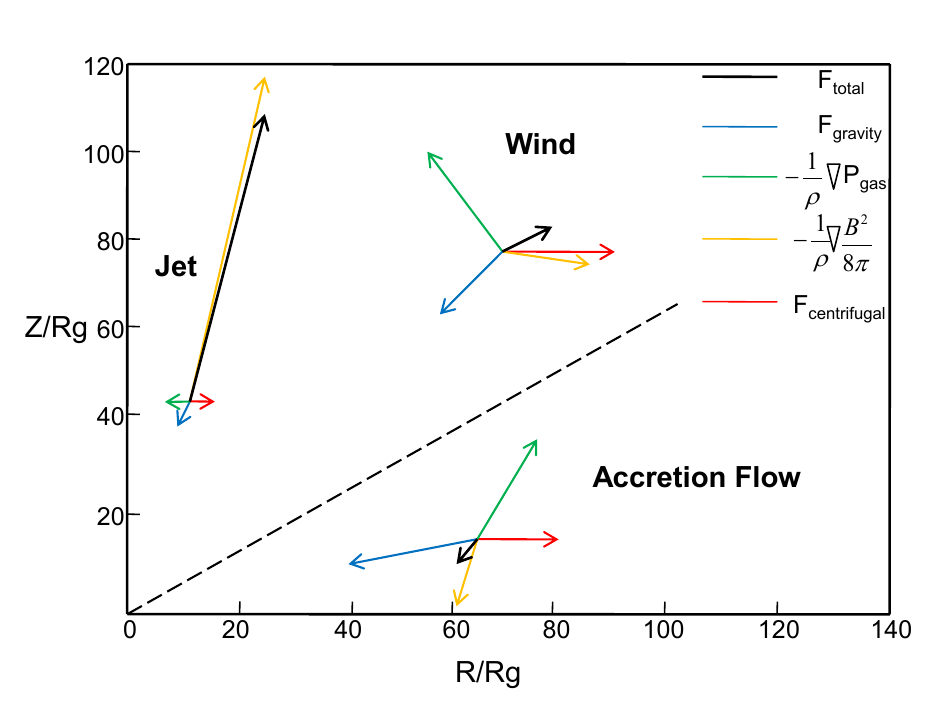}}
\caption{Analysis of forces at three representative locations: jet, wind, and disk. Arrows indicate direction, and lengths represent force magnitude. Adapted from Yuan et al. \cite{Yuan:2015}}
\label{fig:5}      
\end{figure}

Figure \ref{fig:5} illustrates the force analysis of the accretion flow, displaying gravity, gas pressure gradient, Lorentz force, and centrifugal force at typical locations. It is evident that the wind is primarily driven by centrifugal force and the magnetic pressure gradient. Additionally, the existence of a "disk-jet" was also discovered in their simulation. For more detailed information, please refer to Yuan et al. \cite{Yuan:2015}.

Yuan et al.'s work \cite{Yuan:2015} focused on the standard and normal evolution (SANE) case \cite{Narayan:2012} without considering black hole spin. However, strong magnetic field environments, such as magnetically arrested disks (MAD) \cite{Narayan:2003}, may exist in the real Universe. Recent studies on M87 and the SMBH at the center of the Milky Way have indicated that their accretion flows may be in a MAD state \cite{EHTC:2021, Yuan:2022, EHTC:2024}. 

Sadowski et al. \cite{Sadowski:2013} have examined high-spin black holes and MAD situations. However, they still used the traditional streamline method to estimate the real outflow, resulting in a very weak wind mass flux. As shown in Yuan et al.'s work, they could only provide a lower limit on the mass flux of the wind.

To address this limitation, Yang et al. \cite{Yang:2021} extended Yuan et al.'s approach using the trajectory method to study the effects of magnetic fields and black hole spin on the wind. They utilized the Athena++ code \cite{White:2016, Stone:2020} to run three 3D GRMHD simulations: MAD00, MAD98, and SANE98, where "MAD98" signifies a MAD accretion flow with a black hole spin of 0.98.

MAD and SANE represent two distinct accretion flow states characterized by different magnetic field configurations. MAD typically begins with a poloidal magnetic loop threading through the gas torus, facilitating rapid accretion. However, as magnetic flux accumulates near the black hole, it can eventually impede further accretion, leading to a saturated state \cite{Narayan:2003, McKinney:2012, Sadowski:2013}.

In contrast, SANE initially comprises multiple magnetic field loops with opposing polarities. This setup promotes frequent magnetic reconnection during the accretion process, maintaining a weaker magnetic field configuration. This prevents excessive accumulation of magnetic flux around the black hole, allowing the accretion flow to remain in the SANE state \cite{Narayan:2012}.

\begin{figure}[htpb]
\centerline{\includegraphics[scale=.5]{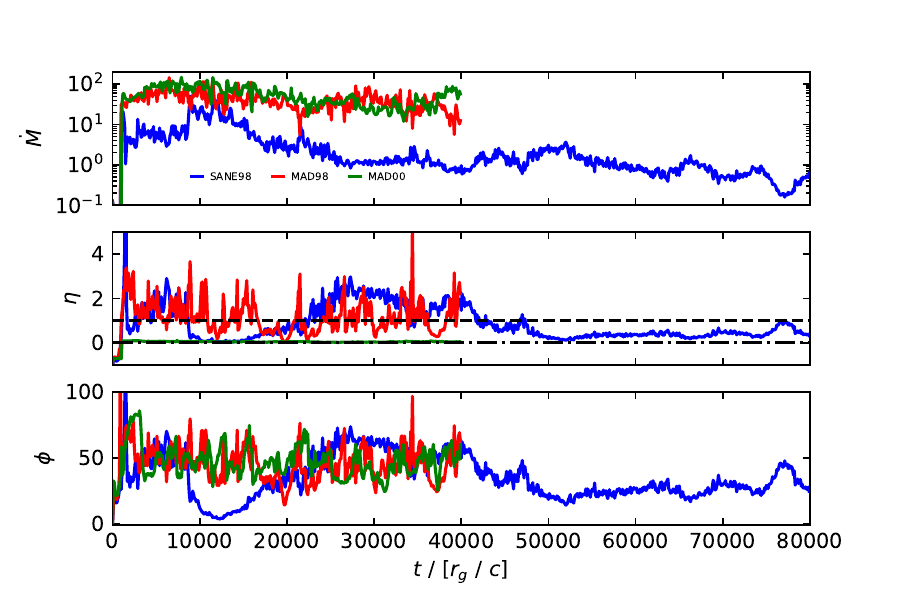}}
\caption{Time evolution of the mass flux, energy extraction efficiency, and dimensionless magnetic flux $\phi$. The blue, red, and green represent the SANE98, MAD98, and MAD00, respectively. Adapted from Yang et al\cite{Yang:2021}.}
\label{fig:6}      
\end{figure}

The determination of whether the accretion flow transitions into the MAD or SANE state is typically based on the normalized averaged magnetic flux $\phi$ threading the hemisphere of the black hole event horizon \cite{Tchekhovskoy:2011, Narayan:2012, Yang:2021}.
\begin{eqnarray}
\phi (t)= \frac{1}{2\sqrt{\dot{M}_{\rm BH}}} \int_{\theta} \int_{\varphi} |B^{r}(r_{\rm H},\,t)| \sqrt{-g}d\theta d\varphi,
\label{eq:12}
\end{eqnarray}

where $ B^{r}$ denotes the radial component of the magnetic field, $r_{\rm H} = r_{\rm g} + \sqrt{1 - a^2} r_{\rm g} $ is the radius of the event horizon, $ a $ is the dimensionless spin parameter of the black hole, and $g$ represents the determinant of the metric. The accretion flow transitions to the MAD state once the normalized averaged magnetic flux $ \phi $ exceeds a critical value of approximately $50$ \cite{Tchekhovskoy:2011}. Figure \ref{fig:6} illustrates the evolution of $\phi $ over time. The green and red solid lines depict the accretion flow in the MAD state, while the solid blue line, appearing after $ t \approx 50,000 \, r_{\rm g}/c $, indicates the accretion flow in the SANE state. 

The plot shows that $ \phi $ exhibits significant variation over time in the MAD state, with magnitudes of change much larger compared to the relatively stable and low values observed in the SANE state.

The presence of black hole spin can lead to the generation of relativistic outflows, known as jets, driven by magnetic fields. When a large-scale poloidal magnetic field crosses the ergosphere and threads the event horizon of a spinning black hole, it can extract rotational energy from the black hole, producing what is known as a Blandford-Znajek (BZ) jet \cite{Blandford:1977}. The power of this jet is described by the BZ model \cite{Blandford:1977, Tchekhovskoy:2011}:

\begin{eqnarray}
P_{\rm BZ} = \frac{\kappa}{4\pi c} \Phi_{\rm H}^2 \Omega_{\rm H}^2,
\label{eq:13}
\end{eqnarray}

where $ \kappa$ is a numerical coefficient, typically around $ 0.5$.  $\Phi_{\rm H} $ represents the magnetic flux threading the event horizon, and $\Omega_{\rm H} = \frac{ac}{4r_{\rm H}} $ denotes the angular velocity of the event horizon. Equation \ref{eq:13} illustrates that the energy of the BZ jet increases with both the magnetic flux threading the event horizon and the spin of the black hole.

Winds and jets coexist in real outflows, necessitating a method to distinguish between them. Some works define the jet region as where the magnetization parameter $\sigma (= B^2/\rho) > 1$ \cite{McKinney:2012}. Alternatively, Yang et al.\cite{Yang:2021} use magnetic field lines to delineate jets, defining them as regions where magnetic field lines connect to the ergosphere of the black hole, with the outer boundary anchored at the ergosphere with $\theta = 90^{\circ}$. Consequently, the outflow beyond the jet region is classified as wind. This distinction is justified by Yang et al.'s study of the M87 jet, demonstrating that it is driven by the BZ mechanism \cite{Yang:2024}. They modeled the BZ jet's image and compared it with observed radio images of the M87 jet, showing that the BZ jet accurately reproduces its morphology, including its length and limb-brightening.

The mass flux of wind and jet can be determined using the trajectory method and Equation \ref{eq:08}. By integrating these methods with the findings from Yuan et al. \cite{Yuan:2015} on SANE00, we can establish radial profiles of the inflow accretion rate for SANE00, SANE98, MAD00, and MAD98 at various distances beyond several $ r_{\rm g} $.
\begin{eqnarray}
\dot{M}_{\rm in }= (\frac{r}{4.8 r_{\rm s}})^{0.54} \dot{M}_{\rm BH},
\label{eq:14}
\end{eqnarray}

\begin{eqnarray}
\dot{M}_{\rm in }= (\frac{r}{0.7 r_{\rm s}})^{0.91} \dot{M}_{\rm BH},
\label{eq:15}
\end{eqnarray}

\begin{eqnarray}
\dot{M}_{\rm in }= (\frac{r}{2.8 r_{\rm s}})^{0.18} \dot{M}_{\rm BH},
\label{eq:16}
\end{eqnarray}

\begin{eqnarray}
\dot{M}_{\rm in }= (\frac{r}{2.5 r_{\rm s}})^{0.42} \dot{M}_{\rm BH},
\label{eq:17}
\end{eqnarray}
Similarly, the radial profiles of the mass flux of wind are, 
\begin{eqnarray}
\dot{M}_{\rm wind }= (\frac{r}{20 r_{\rm s}})^{1.0} \dot{M}_{\rm BH},
\label{eq:18}
\end{eqnarray}

\begin{eqnarray}
\dot{M}_{\rm wind }= (\frac{r}{15 r_{\rm s}})^{1.16} \dot{M}_{\rm BH},
\label{eq:19}
\end{eqnarray}

\begin{eqnarray}
\dot{M}_{\rm wind }= (\frac{r}{55 r_{\rm s}})^{1.54} \dot{M}_{\rm BH},
\label{eq:20}
\end{eqnarray}

\begin{eqnarray}
\dot{M}_{\rm wind }= (\frac{r}{30 r_{\rm s}})^{1.26} \dot{M}_{\rm BH},
\label{eq:21}
\end{eqnarray}
respectively. By comparing SANE00 and MAD00, it is shown that near the black hole where the magnetic field is stronger, the wind is weaker. This phenomenon is likely due to the role of turbulence in wind generation, with SANE turbulence attributed to magneto-rotational instability (MRI), whereas in MAD, MRI is suppressed leading to the absence of significant turbulence \cite{Narayan:2003, McKinney:2012}. However, in regions farther from the black hole where the magnetic field strength diminishes, both SANE and MAD exhibit MRI which aids in wind production. Consequently, the power-law index of MAD00 (1.54) exceeds that of SANE00 (1.0). Similar trends are evident when comparing SANE98 and MAD98.

When comparing SAN00 with SANE98, it becomes evident that at smaller radii, higher spin enhances wind flux. This pattern is also observed in the comparison of MAD00 and MAD98, indicating that spin plays a crucial role in wind generation. Notably, the power-law index from Equation \ref{eq:18} to Equation \ref{eq:21} exceeds that from Equation \ref{eq:14} to Equation \ref{eq:17}, suggesting that beyond specific radii ($214\,r_{\rm g}$ for SANE00, $10^{6}\,r_{\rm g}$ for SANE98, $164\,r_{\rm g}$ for MAD00, and $210\,r_{\rm g}$ for MAD98), the mass flux of the outflow surpasses that of the inflow, which is unphysical. Therefore, the mass flux of the wind equals the inflow at these radii. Furthermore, Equation \ref{eq:19} should only be applied within the Bondi radius, the outer limit of the hot accretion flow, typically located at $10^{5}\, r_{\rm g}$. 

As the magnetic field weakens at large radii, the wind velocity is primarily governed by the poloidal components \cite{Cui:2020a, Cui:2020b}. Therefore, particular attention is given to the poloidal velocity of the wind in these cases. For SANE00, MAD00, SANE98, and MAD98,
\begin{eqnarray}
v_{\rm p }(r)= 0.21\,v_{\rm k}(r),
\label{eq:22}
\end{eqnarray}
\begin{eqnarray}
v_{\rm p }(r)= 0.66\,v_{\rm k}(r),
\label{eq:23}
\end{eqnarray}
\begin{eqnarray}
v_{\rm p }(r)= 0.24\,v_{\rm k}(r),
\label{eq:24}
\end{eqnarray}
\begin{eqnarray}
v_{\rm p }(r)= 0.64\,v_{\rm k}(r),
\label{eq:25}
\end{eqnarray}
respectively. 
In the MAD model, the poloidal velocity of the wind is significantly higher than that in the SANE model. This disparity arises because the Poynting flux of the wind is converted into kinetic energy, consistent with findings by Yuan et al. \cite{Yuan:2015} indicating that the Lorentz force plays a major role in accelerating the wind. The influence of spin on wind velocity is relatively minor, as the wind originates in regions at large radii where relativistic effects due to spin are less pronounced. Additionally, it has been noted that the rotational velocity of the MAD wind is much lower than the Keplerian velocity, which can be attributed to the strong outward magnetic pressure gradient force prevalent in the MAD scenario.

We can obtain the radial profiles of the flux of the momentum and kinetic energy of the wind and jet by the following formula,
\begin{eqnarray}
\dot{E}_{\rm jet (wind) }= \frac{1}{2}\int \gamma \rho(r,\theta,\varphi) v_{r}^{3}(r,\theta,\varphi)\times (r^2+a^2\cos{\theta}^{2})\sin{\theta}d\theta d\varphi,
\label{eq:26}
\end{eqnarray}

\begin{eqnarray}
\dot{P}_{\rm jet (wind) }= \frac{1}{2}\int \gamma \rho(r,\theta,\varphi) v_{r}^{2}(r,\theta,\varphi)\times (r^2+a^2\cos{\theta}^{2})\sin{\theta}d\theta d\varphi,
\label{eq:27}
\end{eqnarray}
Here $\gamma=1/\sqrt{1-v_{r}^{2}}$, $v_{r}$ represents the radial three-velocity component in the Locally Non-Rotating Frame (LNRF). Since $v_{\theta}\ll v_{r}$, only $v_{r}$ is considered here. Further details can be found in \cite{Yang:2021}. The results indicate that both black hole spin and magnetic field enhance the momentum and kinetic energy fluxes of the wind and jet. For non-spinning black holes, Yuan et al. \cite{Yuan:2015} studied SANE00 and found that
\begin{eqnarray}
\dot{E}_{\rm wind} \approx  \frac{1}{1000} \,\dot{M}_{\rm BH}c^2,
\label{eq:28}
\end{eqnarray}

Its energy flux corresponds closely with the requirements for large-scale AGN feedback simulations \cite{Ciotti:2010, Gaspari:2012}. Moreover, to be consistent with observations of isolated galaxies and galaxy clusters, the necessary "mechanical feedback efficiency" should fall within the range of approximately $10^3-10^4$. Equation \ref{eq:28} naturally explains this required efficiency. Comparing the momentum and kinetic energy of the "disk jet" with that of the wind reveals that
\begin{eqnarray}
\dot{P}_{\rm wind} \approx  15 \,\dot{P}_{\rm disk\ jet},
\label{eq:29}
\end{eqnarray}
and, 
\begin{eqnarray}
\dot{E}_{\rm wind} \approx  3 \,\dot{E}_{\rm disk \ jet},
\label{eq:30}
\end{eqnarray}

Whether considering momentum or kinetic energy, the flux of the wind is significantly greater than that of the disk jet. This finding is consistent for a non-spinning black hole, where the density of the disk jet is very low, with $\dot{M}_{\rm jet}\approx1/20\,\dot{M}_{\rm wind}$, and velocities are only sub-relativistic, around $\sim0.3 - 0.4\,c$. These results imply that the wind plays a more crucial role than the jet in AGN feedback. However, it is important to note that the black hole discussed here is non-spinning.

When a black hole has spin, a much more powerful BZ jet emerges. Its velocity is relativistic, reaching up to $0.9\,c$ even in the inner region. This jet carries significant energy, capable of extending to galactic scales, such as in M87, where it spans about 6,500 light-years. Due to the immense power of such jets, most AGN feedback studies focus solely on the jet, often assuming that the wind is insignificant. However, Yang et al. \cite{Yang:2021}'s research on high-spin black holes revealed surprising findings. They compared the momentum and kinetic energy fluxes of MAD98 and SANE98 at $r=200\,r_{\rm g}$. For SANE98, there are,
\begin{eqnarray}
\dot{P}_{\rm wind} \approx  1.5 \,\dot{P}_{\rm jet},
\label{eq:31}
\end{eqnarray}
and, 
\begin{eqnarray}
\dot{E}_{\rm jet} \approx  3.5 \,\dot{E}_{\rm wind},
\label{eq:32}
\end{eqnarray}
For MAD98, there are

\begin{eqnarray}
\dot{P}_{\rm wind} \approx   \,\dot{P}_{\rm jet},
\label{eq:33}
\end{eqnarray}
and, 
\begin{eqnarray}
\dot{E}_{\rm jet} \approx  4 \,\dot{E}_{\rm wind},
\label{eq:34}
\end{eqnarray}

These results show that even for rapidly rotating black holes, where the jet energy is the strongest, the kinetic flux of the jet is only about four times larger than that of the wind, whether in the MAD or SANE model. When considering the total energy flux (the sum of kinetic energy, Poynting flux, and enthalpy), the flux of the jet is only seven and ten times larger than that of the wind for SANE and MAD98, respectively. Moreover, the momentum of the wind exceeds that of the jet. In AGN feedback studies, momentum feedback may play a more critical role than energy feedback in several key aspects, such as the growth of black hole mass, AGN mass accretion rate, and suppression of the star formation rate in the central region of the galaxy \cite{Ostriker:2010}. 

Additionally, the wind's larger opening angle compared to the jet allows it to more effectively deposit energy into the interstellar medium. These findings suggest that wind plays a more significant role in AGN feedback than jets and should be considered in AGN feedback models rather than focusing solely on jets, as has been common in most previous studies.

However, due to the limitations of GRMHD simulations, these results are only applicable at $r=200\,r_{\rm g}$, making it uncertain if they hold at larger radii. Nonetheless, from a momentum perspective, the findings indicate that the momentum of the wind continues to grow as it propagates outward, whereas the momentum of the jet reaches saturation.

Equations \ref{eq:18} - \ref{eq:21} show that as the wind moves outward, the newly generated local wind continues to join. This raises the question: will this continuous generation of new wind persist indefinitely, or will it cease beyond a certain radius? To address this, Bu et al. \cite{Bu:2016a, Bu:2016b} conducted a series of 2D HD and MHD simulations to study accretion flow at very large radii. In their simulations, they considered the gravitational potential of both the black hole and the nuclear star cluster. They assumed that the velocity dispersion of the nuclear stars is constant with radius, leading to the potential of the star clusters:
\begin{eqnarray}
\psi_{\rm star}= \sigma^{2}In(r) + C ,
\label{eq:35}
\end{eqnarray}
where $\sigma$ is the velocity dispersion of the stars, and $C$ is a constant. Typical values for $\sigma$ range from $100 - 400 \, \text{km s}^{-1} $ \cite{Kormendy:2013}, thus $R_{\rm A}$, the radius where the gravity of the black hole equals that of the star cluster, can be defined as:
\begin{eqnarray}
R_{\rm A}= \frac{GM_{\rm BH}}{\sigma^{2}} ,
\label{eq:36}
\end{eqnarray}
which is approximately $10^5 - 10^6 \, r_{\rm s}$, roughly equal to the Bondi radius of the accretion flow. For their standard model, the radial simulation range was $0.3 - 40 \, R_{\rm A}$. They found that although the accretion rate decreased inward, trajectory analysis revealed almost no real outflow. Further analysis indicated that irrespective of the magnetic field, the region beyond $R_{\rm A}$ is convectively unstable. These results suggest that the inward decrease in accretion rate here is caused by convection, which is consistent with the CDAF model.

When they moved the simulation area closer to the black hole, changing the radial simulation range to $0.02 - 4 \, R_{\rm A}$ while keeping other parameters unchanged, they found that the accretion flow within $0.4 \, R_{\rm A}$ is convectively stable and exhibits very strong wind, comparable to that described by Equation \ref{eq:18}. This indicates that within a certain radius, strong wind generation persists, but beyond that radius, the dynamics are dominated by convection rather than continuous outflow generation.

The difference between the two simulations lies solely in the radial range considered, which alters the gravitational potential affecting the accretion flow. In the simulation closer to the black hole, the black hole's gravitational potential dominates over that of the star cluster, leading to the generation of new local wind. In the simulation farther from the black hole, the star cluster's gravity gradually surpasses that of the black hole and becomes dominant. Beyond $ r > R_{\rm A} $, the accretion flow becomes convectively unstable, and no new local wind is produced. This instability may result from changes in the gravitational potential, which in turn alter turbulence properties and thus affect wind generation. Since $ R_{\rm A} $ is close to the Bondi radius, no new wind is generated locally outside the Bondi radius. Therefore, Equations \ref{eq:18} - \ref{eq:21} are valid only within the Bondi radius.

Due to technical limitations, simulations typically extend only to about $10^{3} - 10^{4}\,r_{\rm g}$. However, how far can the wind propagate outward once it escapes this boundary? To explore this question, Cui et al. \cite{Cui:2020a, Cui:2020b} investigated the large-scale behavior of winds generated by accretion flows.

They initially studied wind propagation on a large scale using HD simulations, which included analytic and 2D numerical simulations. Their approach considered not only the gravitational potential of the black hole but also that of the host galaxy. They employed a small-scale hot accretion flow \cite{Yuan:2015} as the inner boundary conditions, where they found:
\begin{eqnarray}
v_{r} \approx v_{r0},
\label{eq:37}
\end{eqnarray}
with $v_{r0}$ representing the wind velocity at the boundary ($r_{0} \equiv 10^{3}\,r_{\rm g}$). Equation \ref{eq:37} illustrates that the wind velocity remains nearly constant as it propagates outward, as the increased gravitational potential is almost offset by enthalpy and rotational energy. They also found that the gravitational potential of the galaxy did not significantly alter the radial velocity of the wind, likely due to the high initial velocity of the wind from the small-scale source. They concluded that the wind can extend to the scale of the galaxy, though this was considered an upper limit without accounting for the interstellar medium (ISM).

They then examined the role of magnetic fields, employing 1D MHD simulations for simplicity. Despite this simplification, they maintained the wind properties from Yuan et al. \cite{Yuan:2015} at the inner boundary. Their findings indicated that $v_{p} \propto {\rm const.}$, meaning the poloidal velocity of the wind remains largely unchanged as it propagates outward. Results from the MHD simulations mirrored those from HD simulations: in cases of weak magnetization, thermal pressure, and magnetic forces jointly accelerated the wind, while in highly magnetized scenarios, magnetic forces dominated and accelerated the wind primarily. Higher magnetization led to increased terminal velocities, with terminal wind velocities reaching approximately $0.016\,c$ for the fiducial model parameters.

It is important to note that their MHD simulations did not incorporate the gravitational potential of the galaxy, despite HD simulations showing minimal impact on wind velocities. Overall, their work demonstrated that small-scale winds originating from accretion flows can propagate far beyond the Bondi radius under favorable conditions.

\section{Observations of wind from hot accretion flows}

Theoretical research suggests the presence of outflows in the ADAF model with low mass accretion rates. However, the question remains: are there any corresponding observational discoveries? Due to the high temperature of the wind within the hot accretion flow, where gas is nearly entirely ionized, observation becomes challenging. Nonetheless, recent studies have unveiled the presence of winds in several low-luminosity AGNs, including Sagittarius A* (Sgr A*) \cite{Wang:2013wind, Ma:2019}, M81 \cite{Shi:2021, Shi:2024}, NGC 7213 \cite{Shi:2022, Shi:2024}, M87 \cite{Park:2019}, and M32 \cite{Peng:2020}, among others\cite{Tombesi:2014, Cheung:2016}. Additionally, winds have been observed in low-mass X-ray binaries (LMXBs) during their hard state \cite{Homan:2016}.

Sgr A* stands as the initial radio source detected \cite{Balick:1974} at the center of our Galaxy. Observations have unveiled its mass, estimated to be $(4.1 \pm 0.4) \times 10^{6} M_{\odot}$ \cite{Ghez:2003}, confined within tens of light-hours \cite{Genzel:2010}, confirming its status as a SMBH. Typically, the outer boundary of an accretion disk surrounding a central black hole is described by the Bondi radius\cite{Bondi:1952}. This radius denotes the equilibrium between the thermal expansion energy of the gas and the gravitational potential energy exerted by the black hole. For Sgr A*, $R_{B} \sim 10^{5}R_{s}\approx 0.04\,pc$. The \textit{Chandra X-ray Observatory} furnishes insights into the gas density and temperature near the Bondi radius, enabling estimation of its mass accretion rate at this boundary, $\dot{M}_{B} \sim 10^{-5}M_{\odot}\,\rm{yr}^{-1}$ \cite{Baganoff:2003}. Analysis of Sgr A*'s spectral energy distribution reveals its remarkably low bolometric luminosity, $L_{bol} \sim 10^{36} {\rm erg\, s^{-1}} \sim 2 \times 10^{-9}L_{\rm{Edd}}$. Should the matter's accretion rate into the black hole equate to its Bondi radius accretion rate, its radiation efficiency would be markedly low, at only $10^{-6}$, far below the typical value for a traditional standard thin disk (approximately $10\%$). This discrepancy strongly suggests the presence of a hot accretion flow. Furthermore, its spectral energy distribution starkly differs from that of a standard thin disk, displaying a multi-temperature blackbody spectrum.

Polarization observations of Sgr A* at 227 and 343 GHz using the Submillimeter Array polarimeter revealed a mean rotation measure of $(-5.6\pm 0.7)\times 10^{5} {\rm rad\,m^{-2}}$. This rotation measure sets constraints on the accretion rate within the innermost region ($r\lesssim 100 R_{\rm s}$). For magnetic fields near equipartition, predominantly ordered and radially oriented, the accretion rate is constrained to be less than $2\times 10^{-7}M_{\odot}\,{\rm yr}^{-1}$. Even in scenarios with nearly subequipartition, disordered, or toroidal fields, a lower limit of $2\times 10^{-9}M_{\odot}\,{\rm yr}^{-1}$ is established\cite{Marrone:2007}. Recent work by the Event Horizon Telescope Collaboration\cite{EHTC:2022} indicates an accretion rate near the event horizon of Sgr A* ranging from $(5.2-9.5)\times 10^{-9}M_{\odot}\,{\rm yr}^{-1}$. Notably, this accretion rate near the black hole is considerably smaller than the accretion rate at the Bondi radius, suggesting a decrease in the accretion rate inward. This result aligns with numerical simulations, hinting at the possibility of outflow.

In 2013, Wang et al.\cite{Wang:2013wind} analyzed high-resolution X-ray emission data from 3 mega-seconds of Chandra observations, revealing a significant outflow (wind) emanating from the center of the Milky Way. The quiescent spectrum of Sgr A* exhibits notable emission lines. In their investigation into the origins of these emissions, they initially discounted the stellar coronal hypothesis due to its inability to account for the predicted Fe K$\alpha$ emission. The presence of a very faint H Fe-like K$\alpha$ line suggested either a radiatively inefficient accretion flow (RIAF) characterized by a very low plasma mass fraction at $kT \gtrsim 9$ keV or the existence of a potent outflow at a radius $r \gtrsim 10^{4} R_{\rm s}$. Employing quantitative analyses, they dismissed the non-outflow scenario characterized by a radial plasma density profile of $n \varpropto r^{-3/2+s}$, with $s=0$, as it would overestimate the H-like Fe K$\alpha$ line. Instead, they found that a RIAF with a flat-density profile ($s \sim 1 $) could aptly fit both continuum and line data. This suggests a decrease in mass accretion rate inward, accompanied by the presence of outflow.

Tombesi et al. (2014)\cite{Tombesi:2014} conducted an analysis of Fe K absorption lines observed in 26 radio-loud AGN spectra from \textit{XMM–Newton} and \textit{Suzaku}. They identified ultrafast outflows (UFOs) with a broad range of velocity distributions, spanning from $1000 \, \mathrm{km\,s^{-1}}$ to $0.4\,c$. These outflows exhibited characteristics of high ionization and significant column densities. The inclination angles of the outflows ranged from approximately $10^{\circ}$ to $70^{\circ}$. These findings suggest a plausible origin from a hot accretion flow, particularly given recent observations indicating that low-luminosity AGNs tend to be radio-loud\cite{Nagar:2000, Falcke:2000, Ho:2002}, implying a potential association between radio-loudness and hot accretion flows. Furthermore, observational evidence suggests that cool thin disks typically exhibit weak or absent jets \cite{Yuan:2014}.

Cheung et al. (2016)\cite{Cheung:2016} observed a bisymmetric ionized gas emission feature spanning the entire galaxy in low-luminosity AGN. This feature was found to be aligned with gas velocity gradients, indicating the presence of multi-component gas with varying temperatures and velocities. Their model of centrally driven winds successfully reproduced these observed features. Their results not only supported the existence of winds but also implied that low-luminosity active galactic nuclei possess sufficient energy to generate winds, and heat surrounding cold gas, and potentially suppress star formation.

However, the aforementioned outflow observations were primarily obtained indirectly. In 2021, Shi et al. (2021)\cite{Shi:2021} provided direct evidence for the existence of wind from M81*, one of the nearest SMBHs with a mass of approximately $M_{\rm BH} \approx 10^{7}M_{\odot}$\cite{Devereux:2003}. Despite its very low bolometric luminosity, roughly $L_{\rm bol}=3\times 10^{-5}L_{\rm Edd}$\cite{Nemmen:2014}, M81* exhibits a pronounced radio jet\cite{Bietenholz:2000} and possibly lacks the classic thin accretion disk\cite{Ho:1996, Young:2018}, indicating that it may be powered by a radiatively inefficient, hot accretion flow.

Shi et al. (2021) analyzed the X-ray spectrum of M81* using high-resolution data from the \textit{Chandra} observatory\cite{Shi:2021}. They detected emission lines but no absorption lines, with four prominent emission lines located at energies of 6.40, 6.69, 6.90, and 7.05 keV. These emission lines were interpreted as arising from a pair of collisionally-ionized, optically-thin plasmas—one red-shifted and the other blue-shifted. Treating the 6.90/7.05 keV lines as symmetrically redshifted/blueshifted Fe XXVI lines, they derived a best-fitting absolute line-of-sight velocity of $(2.8 \pm 0.2)\times 10^{3} {\rm km\,s^{-1}}$. Additionally, they determined a best-fitting temperature of $1.3 \times 10^{8}$K based on the relative intensity between Fe XXVI Ly$\alpha$ and Fe XXV K$\alpha$. Such high velocities and temperatures are inconsistent with stellar activities or accretion inflow, suggesting that only winds originating from hot accretion flows could produce such features.

By conducting MHD simulations of the wind launched from the hot accretion flow in M81*, Shi et al. obtained spatial distributions of density, velocity, temperature, and magnetic field. They utilized this information to generate synthetic X-ray spectra, which exhibited two main features consistent with the observed results: a double-peak Fe XXVI Ly$\alpha$ profile and a high Fe XXVI-to-Fe XXV flux ratio.

Conclusively, the observed high-velocity and high-temperature plasma represents the wind originating from the hot accretion flow, marking the first direct observation of such a phenomenon. Shi et al. (2024)\cite{Shi:2024} subsequently identified novel blue-shifted emission lines, hypothesizing that these emanated from circumnuclear gas shock-heated by the hot wind. Their findings provide compelling evidence for energy and momentum feedback from the AGN's hot wind.

\subsection{Application of the wind from hot accretion flows}

An intriguing aspect of Sgr A* is that, despite its current low mass accretion rate of only $\dot{M}_{\rm BH}\sim 10^{-6}\dot{M}_{\rm Edd}$, numerous observations indicate that it was significantly more active in the past few million years, with an accretion rate approximately four orders of magnitude higher than it is now \cite{Totani:2006}. A clear piece of observational evidence supporting this increased activity is the recently discovered \textit{Fermi} Bubbles. Su et al. (2010)\cite{Su:2010}  identified two giant gamma-ray bubbles above and below the Galactic plane using \textit{Fermi}-LAT. In galactic coordinates, each bubble has a height of approximately $40^{\circ}$ and a width of $50^{\circ}$. The surface brightness of these bubbles is roughly uniform with sharp edges. In the $1-100\,{\rm eV}$ band, the total luminosity of the bubbles is $4\times10^{37}\,{\rm erg\,s^{-1}}$, and the total energy of the two bubbles is about $10^{55}-10^{56}\,{\rm erg}$.

To explain the observed \textit{Fermi} Bubbles, several models have been proposed, such as the "jet" model \cite{Guo:2012a, Guo:2012b} and the "quasar outflow" model \cite{Zubovas:2011}. These models typically explain the bubbles' morphology through interactions between a jet or wind (from super-Eddington accretion flow) and the ISM. However, wind from hot accretion flows could also produce such bubbles. Research by Mou et al. \cite{Mou:2014, Mou:2015} demonstrates that wind from a hot accretion flow can adequately explain the observed bubbles' morphology and gamma-ray spectral energy distribution. Although the accretion rate of Sgr A* was $\sim10^{-2}\,\dot{M}_{\rm Edd}$ at that time, it remained in a hot accretion mode. At this accretion rate, the accretion disk typically consists of a hot accretion flow near the black hole and a thin disk further away, with the transition radius $R_{\rm tr}$ being a function of the accretion rate. For $\sim10^{-2}\,\dot{M}_{\rm Edd}$, $R_{\rm tr}$ is approximately $200\,r_{\rm s}$\cite{Yuan:2014}.

The properties of the wind are not arbitrary but are based on numerical simulation results of the hot accretion flow. The wind parameters, including mass flux and velocity, were derived from the results of Yuan et al. \cite{Yuan:2015}, such as Equations \ref{eq:09} and \ref{eq:11}. The wind used in their simulations has a mass flux of about $18\%\dot{M}_{\rm Edd}$ and a velocity of approximately $v_{\rm k}(R_{\rm tr})\sim0.046\,c$, resulting in kinetic energy of the thermal gas in the wind of $10^{42}\,{\rm erg\,s^{-1}}$. They conducted a series of 3D MHD simulations based on these wind parameters. By using the boundary between the shocked ISM and the shocked wind as the edge of the bubbles, they found that the morphology of the bubbles closely matched the observations.

To calculate the gamma-ray spectra, cosmic rays (CRs), mainly CR protons, were injected into the wind at the beginning of the simulation, with about $50\%$ of the wind's power. The thermal gas and CR protons were simulated simultaneously. 
The gamma-ray spectrum, obtained by combining the CR results with the "hadronic" model, was highly consistent with observations. Additionally, their results showed a slight limb-brightened surface brightness, consistent with the observed blurred limb-brightened feature reported by \cite{Ackermann:2014}.

In their work, they ejected the wind isotropically without considering the jet. They posited that the jet would pass through a narrow low-density channel, allowing it to move freely without generating a bubble-like structure \cite{Vernaleo:2006}. Additionally, they calculated the energy transformation efficiency, finding that approximately $60\%$ of the wind's energy is transferred to the ISM, likely due to the wind's large opening angle. These results suggest that we should seriously consider the possibility that cavities and bubbles observed in other environments, such as galaxy clusters, may be formed by winds rather than jets.

\section{Wind from super-Eddington accretion flows}

In this section, we primarily investigate wind originating from super-Eddington accretion flow. Observational data suggest that certain quasars experience super-Eddington accretion phases during their growth\cite{Kelly:2013}, particularly in their early developmental stages\cite{Inayoshi:2020}. The feedback mechanisms of AGN profoundly influence the formation and evolution of their host galaxies \cite{Fabian:2012, Kormendy:2013, Naab:2017}. Although AGNs may only exhibit super-Eddington accretion sporadically, the resulting outflows and radiation remain substantial, potentially exerting significant feedback effects. The outflows from super-Eddington accretion flow impact the observed spectra and have been invoked to explain certain TDEs\cite{Dai:2018, Thomsen:2022}.

Super-Eddington accretion flows, characterized by higher accretion rates, exhibit distinct properties in both their accretion flow and outflow compared to the thin disk model. Consequently, the slim disk model was proposed \cite{Abramowicz:1988}. However, due to its one-dimensional nature, this model failed to provide comprehensive insights into the properties of the outflow. Subsequent analytical studies \cite{King:2016, Cao:2022} demonstrated the existence of winds. For instance, they found that when the disk accretes at its outer edge at moderate Eddington scale rates (up to $\sim 100$), a significant portion of the inflow is expelled as outflow, with only a fraction accreted by the black hole. The outflow typically exhibits slightly subrelativistic speeds, reaching velocities of $\sim 0.1–0.2\,c$ for Eddington accretion factors $\dot{m}_{\rm acc} \sim 10-100$, and around $\sim 1500 \, {\rm km s^{-1}}$ for $\dot{m}_{\rm acc} \sim 10^4$. However, significant advancements emerged from more realistic numerical simulations \cite{Eggum:1988, Ohsuga:2005, Jiang:2014, Sadowski:2014, McKinney:2015, Asahina:2022, Hu:2022, Yang:2023}. These simulations have provided substantial progress in understanding the dynamics of super-Eddington accretion flow and their associated outflows.

The pioneering numerical simulations on radiation hydrodynamics (RHD) of super-Eddington accretion flows were conducted by Eggum et al. (1988)\cite{Eggum:1988}. They observed the emergence of high-velocity outflows near the axis of rotation. Subsequently, Ohsuga et al. (2005)\cite{Ohsuga:2005} extended this seminal work by scaling up the simulation domain and prolonging the runtime to attain a quasi-steady state. In their investigation, they confirmed the presence of radiation-driven wind. Subsequent advancements included the incorporation of the magnetic field in simulations\cite{Ohsuga:2009, Ohsuga:2011}, revealing that radiation pressure exerted at the disk's surface significantly exceeded gravity, thereby driving outflows.

However, it's noteworthy that these simulations relied on the flux-limited diffusion (FLD) method to handle radiation transport. While effective for optically thick accretion disks, the FLD approximation assumes isotropic emission relative to the fluid frame. This approximation is adequate for optically thick regions but may lead to inaccuracies in optically thin coronal regions where radiation tends to be anisotropic. Thus, despite their larger contributions, these simulations may provide only a rough approximation due to the limitations of the FLD method.

Subsequently, Jiang et al. (2014)\cite{Jiang:2014} conducted a comprehensive global three-dimensional radiation magneto-hydrodynamical simulation. By directly solving the time-dependent radiative transfer equations within the Newtonian limit, they identified prominent outflow regions within approximately $45^\circ$ from the rotation axis. These outflows, primarily propelled by radiation forces, were observed to be collimated by magnetic pressure \cite{Ohsuga:2011}. Notably, the wind velocities ranged from about 0.1 to 0.4 times the speed of light. Their findings indicated that for a stellar-mass black hole accreting at an Eddington rate of $\sim 220\, L_{\rm{Edd}}/c^2$, radiation-driven outflows predominantly formed near the rotation axis, with the total mass flux of outflow constituting approximately $29\%$ of the net accretion rate.

While the pseudo-Newtonian potential can adequately represent black holes without spin to some extent, the influence of general relativity (GR) effects on the spectrum cannot be neglected, particularly considering that a significant portion of radiation emanates from the inner regions. Additionally, GR effects must be considered in discussions concerning jet formation and the impact of black hole spin.

To address these complexities, a series of General Relativistic Radiation Magnetohydrodynamics (GRRMHD) codes, such as KORAL (Sadowski et al., 2013)\cite{Sadowski:2013} and HARMRAD (McKinney et al., 2014)\cite{McKinney:2014}, were developed. These codes employed the M1 closure to effectively handle radiative momentum equations, offering a suitable approximation for both optically thick and optically thin regions under GR. Sadowski et al. (2015)\cite{Sadowski:2015a} introduced a novel mean-field magnetic dynamo into KORAL, enabling 2D axisymmetric simulations to achieve a larger steady-state radius. They observed that for non-spinning stellar-mass black holes ($10 \,M_{\odot}$), the relative strength of the wind mass flux remained relatively constant across varying accretion rates ($2< \dot{M} / \dot{M}_{\rm Edd}<500$), with the outflow rate approximately matching the net accretion rate at $40\,r_{\rm g}$.

Moreover, Utsumi et al. (2022)\cite{Utsumi:2022} investigated the impact of black hole spin on the outflow of super-Eddington accretion flows through a 2D axisymmetric GR-RMHD simulation. They observed the ejection of disk material from the disk surface, primarily driven by radiative forces. However, at very high spins, the magnetic force also became significant. Notably, even in cases where the accretion flow exhibited a SANE rather than MAD configuration, the magnetic field's influence became pronounced at high spin values.

Given the pivotal role of radiation, the ratio of radiation pressure to gas pressure differs significantly between the accretion flows of stellar-mass black holes and SMBHs, even at the same Eddington accretion rate. To investigate this, Jiang et al. (2019)\cite{Jiang:2019} conducted 3D radiation magnetohydrodynamics (RMHD) simulations of super-Eddington accretion flows around SMBHs. Their findings revealed that the outflow mass flux ranged from $15\%$ to $50\%$ of the net accretion rate, with velocities in the funnel region reaching $0.3 - 0.4c$. These results underscored that, for equivalent accretion rates, AGNs exhibit larger outflow mass fluxes compared to stellar-mass black holes.

Additionally, Sadowski et al. (2015)\cite{Sadowski:2015b} investigated mildly SMBHs ($3\times 10^5\, M_{\odot}$) through 2D GRRMHD simulations at $45\, \dot{M}_{\rm Edd}$. They identified high-speed outflows, with velocities increasing gradually from the disk to the axis, reaching speeds of up to $0.5c$. The density-weighted velocity averaged approximately $0.3c$. These high-speed outflows in the funnel regions were primarily radiation-driven, attributed to the absence of black hole spin and the presence of minimal magnetic fields. For higher accretion rates ($310\, \dot{M}_{\rm Edd}$ and $4800\, \dot{M}_{\rm Edd}$), the termination velocity decreased to $0.3c$, possibly due to increased mass loading in the funnel region.

While numerous simulations have hinted at the existence of winds, few studies have comprehensively examined and compared them to jets. Some research has suggested that the inflow rate of the accretion flow follows a power-law distribution $\dot{M}_{\rm in}\propto r^{s}$, with an index $s$ typically ranging from $\sim 0.4$ to $1.0$ \cite{Yang:2014, Hu:2022, Yang:2023}. This implies that only a fraction of matter is accreted into the black hole. However, accurately determining the true outflow is challenging due to the turbulent nature of the accretion flow. Conventionally, the $Be$ parameter has been employed to identify outflows, albeit with limitations. It tends to underestimate the mass flux of the outflow and can only provide a lower limit. Studies on both sub-Eddington and super-Eddington accretion flows have revealed that the $Be$ parameter is not a robust criterion for distinguishing outflows \cite{Yuan:2015, Yoshioka:2022}.

Yang et al. (2023)\cite{Yang:2023} employed the ``virtual particle trajectory" method to analyze super-Eddington accretion flow at the MAD state in 3D GRRMHD. This method, compared to traditional streamline approaches, offers improved discernment between real outflows (wind and jet) and turbulent flows. Moreover, the definition of jets and winds has been a subject of ambiguity. Jets are typically delineated by magnetization parameters ($\sigma>1$) \cite{McKinney:2012} and Bernoulli constant ($Be>0.05$) \cite{McKinney:2014,Sadowski:2015b}. In Yang et al.'s study (2021)\cite{Yang:2021}, the outermost magnetic field line connected to the black hole event horizon (away from the rotation axis) was deemed the boundary of the BZ jet. Subsequent research corroborated that the observed jet indeed corresponded to a BZ jet (Yang et al., 2024)\cite{Yang:2024}.

Furthermore, the results of Yang et al. (2023)\cite{Yang:2023} revealed that the wind mass flux follows a power-law distribution 
\begin{eqnarray}
\dot{M}_{\rm wind} =(\frac{r}{45 r_{\rm s}})^{0.83} \dot{M}_{\rm BH},
\label{eq:38}
\end{eqnarray}
and the mass-flux-weighted poloidal velocity of approximately $0.15\, c$. The momentum and kinetic energy flux of the wind at $r=200\,r_{\rm g}$ are 
\begin{eqnarray}
\dot{P}_{\rm wind} \approx 0.15\,\dot{M}_{\rm BH}c,
\label{eq:39}
\end{eqnarray}\begin{eqnarray}
\dot{E}_{\rm wind}  \approx 0.01\,\dot{M}_{\rm BH}c^2,
\label{eq:40}
\end{eqnarray}
respectively. Additionally, a comparison of wind and jet properties indicated that the momentum flux of wind exceeds that of jets ($\dot{P}_{\rm wind}\approx1.2\,\dot{P}_{\rm jet}$), while the total energy flux of jets is only about three times greater than that of wind ($\dot{E}_{\rm jet}\approx2.5\,\dot{E}_{\rm wind}$). In MAD disks, the primary acceleration force for both wind and jets is the Lorentz force, with radiation and centrifugal forces serving as secondary accelerators.

Recently, several GRRMHD codes have been developed, incorporating magnetic fields, GR effects, and more accurate radiation treatments (Ryan et al., 2015; Asahina et al., 2022; White et al., 2023; Liska et al., 2023)\cite{Ryan:2015, Asahina:2022, White:2023, Liska:2023}. Asahina et al. (2022)\cite{Asahina:2022} conducted a comparative study between methods directly solving radiative transfer equations to obtain the Eddington tensor and employing the M1 closure in GRRMHD simulations. Their findings revealed the emergence of accretion disks and disk winds in both scenarios, with no significant disparities observed in accretion rates, outflow rates, or luminosity. However, discrepancies were noted in the radiation field near the rotation axis, with the M1 method amplifying the radial component of the radiation flux.

Observations of winds from Super-Eddington AGNs remain limited. While investigating the X-ray spectrum of the narrow-line Seyfert 1 galaxy I Zwicky 1, Ding et al. (2022) \cite{Ding:2022} discovered multiple variable blue-shifted absorption features attributed to outflowing material along the line of sight. These features are believed to originate from ionization winds and ultrafast outflows. Notably, the disk ionization parameter is correlated with the outflow velocity, following a relationship of $\xi \propto v_{\rm wind}^{3.24}$, indicative of super-Eddington winds. The wind velocity ranges from approximately $0.1$ to $0.3\,c$.

In a separate study, Vietri et al. (2022)\cite{Vietri:2022} identified three Broad Absorption Line (BAL) UFOs ($\sim 0.13-0.18\,c$) during observations of hyper-luminous quasars. These UFOs exhibited variations in intensity and shape, along with instances of disappearance and reappearance at different temporal intervals.

\begin{acknowledgement}
This work was supported in part by the Natural Science Foundation of China (grants 12133008, 12192220, and 12192223). H.Y. acknowledges the support from the National Key R\&D Program of China (No. 2023YFE0101200), the National Natural Science Foundation of China (Grant No. 12273022), and the Shanghai Municipality orientation program of Basic Research for International Scientists (grant no. 22JC1410600). 
\end{acknowledgement}




\end{document}